\documentclass[letterpaper]{jpconf}
\usepackage{graphicx}

\begin{document}
\title{``Cosmic Rays" from Quark Matter}

\author{Kyle Lawson}

\address{University of British Columbia}

\begin{abstract}
I describes a dark matter candidate based in \textsc{qcd} physics in 
which the dark matter is composed of macroscopically large ``nuggets" of 
quark and anti-quark matter. These objects may have a  
sufficiently massive low number density  to  avoid constraints from direct detection 
searches. Though not ``baryonic" in the conventional 
sense quark matter is strongly interacting and will produce a clear signal 
in ground based detectors. As the prospects of detecting these objects 
are mainly limited by the detector cross-section  
large scale cosmic ray detectors are a promising search 
platform. To this end I describe  the basic properties of the air shower induced 
by the passage of a quark nugget through the earth's atmosphere. It will be shown 
that this shower is similar in several important ways to the shower induced by a 
single ultrahigh energy cosmic ray.
\end{abstract}

\section{Introduction}
It is now well established that our galaxy contains roughly five times more dark 
matter than it does visible matter. While the large scale properties of the dark matter 
are well described in $\Lambda$CDM cosmology there is at present 
no generally accepted microscopic model. In 
the absence of a  suitable particle within the standard model the majority 
of dark matter candidates represent new fundamental particles, the properties of 
which are poorly constrained. An alternatively class of models exists 
in which the dark matter is a composite object composed 
of the known standard model quarks in a novel phase such as 
strange quark matter or a colour superconducting phase. In these models the 
dark matter forms in the early universe (most likely at the \textsc{qcd} phase transition ) 
and remains ``dark" due purely to its small cross-section to mass ratio. To avoid 
observational constraints from both cosmology and direct searches these object
mass carry a baryon number of at least $10^{25}$ corresponding to a minimum mass 
of roughly a gram. At nuclear densities a nugget of this mass has a physical 
cross section of $10^{-5}$cm. 

Here I will focus on a particular composite dark matter scenario 
\cite{Zhitnitsky:2002qa} motivated by the 
seemingly unrelated problem of  matter/antimatter asymmetry. While microscopic
physical laws treat matter and antimatter identically it is observed that the 
universe contains almost exclusively matter. The mechanism responsible for generating this 
baryon excess is not presently known. In the scenario considered here dark
matter forms out of ordinary quarks and antiquarks at the \textsc{qcd} phase transition. 
During the phase transition $\mathcal{CP}$ violation results in a preferential formation of 
nuggets of antimatter and thus the visible matter (which is not compressed into 
nuggets) is left with a matter excess. The observed matter to dark matter 
ratio requires that nuggets and anti-nuggets occur in a roughly two to three ratio. 

A full description of the formation dynamics requires a careful treatment of strongly 
coupled \textsc{qcd} near the phase transition which is not tractable beyond very 
basic arguments at this time. Constraining this models is much easier by taking 
an observational perspective, once formed and cooled below the phase transition 
the nuggets are well described by known many-body and nuclear 
physics allowing quite specific predictions to be made. 

The small number density of quark nuggets required to explain the observed 
dark matter mass density renders them invisible to the majority of  ground based 
searches which are instead targeted towards the detection of \textsc{wimps} 
with masses many orders of magnitude smaller. Large scale cosmic ray detectors, 
such as the Pierre Auger Observatory,  are among the only experiments with a 
sufficiently large cross-section to impose constraints on quark nuggets within 
the mass range on interest. From a combination of theoretical and 
observational considerations the nuggets are believed to carry a baryon 
number in the range from $10^{24}$ up to $10^{30}$ and have a density 
similar to that of ordinary nuclear matter.

\section{Flux}
Baring any significant deviations from galactic averages the dark matter 
has a local density of roughly 1.5 GeV/cm$^3$ and a velocity distribution centred  
on the virial speed of $\sim 300$km/s. This implies that if the nuggets carry a 
baryonic charge $B$ the observed flux at the earth's surface will be 
\begin{equation}
\label{eq:flux}
\Phi \approx B^{-1} 10^{25} s^{-1} km^{-2}
\end{equation}
Here it is assumed that only antimatter nuggets will produce observable 
consequences in a cosmic ray detector and that the mass per baryon of 
quark matter is not substantially different from than in nuclear matter. 
In the case of a matter nugget 
only kinetic energy is deposited in the atmosphere. The collisions 
are not sufficiently energetic to produce significant air fluorescence or 
eject substantial numbers of charged particles from the tightly bound nugget. 

\section{Air Shower Properties}
An antimatter nugget striking the earth's atmosphere will release a 
significant amount of energy primarily through nuclear annihilations 
Much of this energy thermalizes within the nugget however some 
fraction will be released into the atmosphere in the form of high energy 
photons or muons emitted by the various nuclear reactions. The 
resulting air shower is phenomenologically similar to that induced by a single 
ultra high energy cosmic ray. This is due to the simple fact that the shower 
is triggered by a large number of hadronic interactions that cascade 
down to the \textsc{qcd} scale. The development of the resulting air shower 
is determined by the microscopic interactions between the nugget and the 
molecules of the atmosphere. Here I will attempt to give a qualitative 
overview of these interactions.

When an atmospheric molecule strikes the quark nugget it must first pass 
through a layer of electromagnetically bound positrons present in all known forms 
of quark matter and referred to as the electrosphere. Here the electrons of 
the molecule annihilate producing gamma rays. The remaining fully ionized 
nuclei then penetrate into the quark matter where they annihilate. This 
annihilation results in jets of primarily light mesons which stream away from the 
annihilation site. As the collision is non-relativistic half of the energy will be 
directed back towards the nugget's 
surface while the remainder thermalizes within the nugget's interior.   
Much of the energy carried by the surface directed jets will cascade down to the 
positrons bound to the quark matter which are the lightest available modes. 
These positrons are ejected from the surface and slowed by the strong surface 
electric fields. They rapidly lose their kinetic energy which is emitted as a 
hard x-ray bremstralung spectrum. Some fraction of the mesons originally 
produced in the annihilation will also reach the quark matter surface. Being strongly 
interacting they will be unable to escape from the quark matter and will scatter 
off the surface. This process can result in the emission of a muon through 
induced beta decay of the meson. Bremstralung emission from these muons is 
less efficient than from similarly produced positrons and they will be able to escape 
from the nugget. In fact these are likely to be the only charged particles able to 
travel a large distance into the atmosphere and, as such, will dominate the 
air shower.

While the temperature remains low the number of muons produced will 
be proportional to the density of air that the nugget is traveling through. As 
stated above positron excitations are favoured over muon production. The 
number of muons produced per annihilation will depend on the exact details 
of the quark matter surface but is generically less than one. The muon energy 
spectrum is also dependent on the exact phase of quark matter of which the 
nugget is composed. The majority of muons will leave the nugget with energies 
near the plasma frequency of the quark matter, over a wide range of quark 
matter phases this is predicted to be at the tens of MeV level. Muon energies 
may be as high as a few GeV but, as they arise from low energy proton 
annihilations will certainly not exceed this level.

\section{Shower maximum}
As discussed above a significant fraction of the annihilation energy 
is thermalized within the nugget. The thermal evolution of the nugget is 
thus determined by the balance between energy deposition 
through annihilation and thermal radiation from the electrosphere. 
The thermal emission spectrum from the positrons of the electrosphere has 
been previously calculated for independent reasons. When applied in 
this context it allows for the calculation of temperature as a function of 
surrounding matter density. 

At low temperatures electron positron annihilation happens primarily through 
an intermediate positronium resonance. While this process is operating 
annihilation proceeds rapidly and the incoming atmospheric molecules are 
ionized almost immediately. However, once the temperature exceeds the positonium 
formation threshold the resonance channel is no longer available and annihilation must occur  
through the much slower direct annihilation process. At low momenta 
elastic scattering through soft photon exchange is preferred to annihilation. 
As such, an incident molecule experiences a thermal pressure as it moves through 
the low density outer layers of the electrosphere. This pressure grows proportionally 
to the temperature, which is in turn established by the annihilation rate. This feedback results 
in an equilibrium rate at which matter can be feed onto the quark surface. 
The equilibrium temperature generically lies between the positronium energy scale 
($m_e\alpha \sim$ keV) and the electron mass ($m_e = 511keV$) where the annihilation 
and elastic scattering amplitudes become comparable. Numerically the rapid fall off of 
positronium formation with energy favours the lower end of this energy range and equilibrium 
is reached in the few tens of keV range. For a quark nugget moving at typical 
galactic velocities this occurs at a height in the atmosphere near ten kilometers. 
Beyond this atmospheric depth the shower will no longer increase in size. 

\section{Conclusion}
Large area cosmic ray detectors have the ability to impose 
constraints on heavy composite dark matter models. In the class of 
these models where the dark matter has an antimatter component 
hadronic interactions with the atmosphere will induce an extensive 
air shower similar to that produced by a single high energy cosmic ray. 
The computation of the full properties of the air shower will require 
numerical simulations similar to those required for standard cosmic ray 
shower but are, in principle, fully tractable as they depend only on 
well understood nuclear interactions at energy scales that are both 
theoretically and experimentally accessible. 

\section*{References}
\bibliography{proc_bib}

\providecommand{\newblock}{}
\begin{thebibliography}{1}
\expandafter\ifx\csname url\endcsname\relax
  \def\url#1{{\tt #1}}\fi
\expandafter\ifx\csname urlprefix\endcsname\relax\def\urlprefix{URL }\fi
\providecommand{\eprint}[2][]{\url{#2}}

\bibitem{Zhitnitsky:2002qa}
Zhitnitsky A~R 2003 {\em JCAP\/} {\bf 10} 010 (\textit{Preprint}
  \eprint{arXiv:hep-ph/0202161})

\end{thebibliography}

\end{document}